\documentclass[prl,letterpaper,twocolumn,nofootinbib,preprintnumbers,superscriptaddress,floatfix]{revtex4-1} 

\pdfpageattr {/Group << /S /Transparency /I true /CS /DeviceRGB>>}

\usepackage[utf8]{inputenc}
\usepackage{newtxtext,newtxmath}

\usepackage{titlesec}
\usepackage{dcolumn}
\usepackage{bm}
\usepackage{amsmath,amssymb,amsfonts}

\usepackage{colortbl}
\usepackage{booktabs}
\usepackage{graphicx}

\usepackage{slashed}
\usepackage[svgnames]{xcolor}
\usepackage[english]{babel}
\usepackage{blindtext}
\usepackage{microtype}

\usepackage[colorlinks=true,linkcolor=blue,urlcolor=blue,citecolor=blue]{hyperref}
\usepackage{ifpdf}

\newcommand{\vect}[1]{\boldsymbol{#1}}
\newcommand{\ph}[1]{\phantom{#1}}

\newcommand{\sh}[1]{\slashed{#1}}

\font\bb=bbmss10 scaled 1200
\def\ident{\mbox{\bb 1}}

\graphicspath{{.}{figures/}} 

\def\hs{\hspace}

\def\no{\nonumber}

\def\lf{\left}
\def\rg{\right}

\makeatletter
\newcommand*\if@single[3]{%
  \setbox0\hbox{${\mathaccent"0362{#1}}^H$}%
  \setbox2\hbox{${\mathaccent"0362{\kern0pt#1}}^H$}%
  \ifdim\ht0=\ht2 #3\else #2\fi
  }
\newcommand*\rel@kern[1]{\kern#1\dimexpr\macc@kerna}
\newcommand*\widebar[1]{\@ifnextchar^{{\wide@bar{#1}{0}}}{\wide@bar{#1}{1}}}
\newcommand*\wide@bar[2]{\if@single{#1}{\wide@bar@{#1}{#2}{1}}{\wide@bar@{#1}{#2}{2}}}
\newcommand*\wide@bar@[3]{%
  \begingroup
  \def\mathaccent##1##2{%
    \if#32 \let\macc@nucleus\first@char \fi
    \setbox\z@\hbox{$\macc@style{\macc@nucleus}_{}$}%
    \setbox\tw@\hbox{$\macc@style{\macc@nucleus}{}_{}$}%
    \dimen@\wd\tw@
    \advance\dimen@-\wd\z@
    \divide\dimen@ 3
    \@tempdima\wd\tw@
    \advance\@tempdima-\scriptspace
    \divide\@tempdima 10
    \advance\dimen@-\@tempdima
    \ifdim\dimen@>\z@ \dimen@0pt\fi
    \rel@kern{0.6}\kern-\dimen@
    \if#31
      \overline{\rel@kern{-0.6}\kern\dimen@\macc@nucleus\rel@kern{0.4}\kern\dimen@}%
      \advance\dimen@0.4\dimexpr\macc@kerna
      \let\final@kern#2%
      \ifdim\dimen@<\z@ \let\final@kern1\fi
      \if\final@kern1 \kern-\dimen@\fi
    \else
      \overline{\rel@kern{-0.6}\kern\dimen@#1}%
    \fi
  }%
  \macc@depth\@ne
  \let\math@bgroup\@empty \let\math@egroup\macc@set@skewchar
  \mathsurround\z@ \frozen@everymath{\mathgroup\macc@group\relax}%
  \macc@set@skewchar\relax
  \let\mathaccentV\macc@nested@a
  \if#31
    \macc@nested@a\relax111{#1}%
  \else
    \def\gobble@till@marker##1\endmarker{}%
    \futurelet\first@char\gobble@till@marker#1\endmarker
    \ifcat\noexpand\first@char A\else
      \def\first@char{}%
    \fi
    \macc@nested@a\relax111{\first@char}%
  \fi
  \endgroup
}
\makeatother

\begin{document}

\title{Distinguishing Quarks and Gluons in Pion and Kaon PDFs}

\author{Kyle D. Bednar}
\affiliation{Center for Nuclear Research, Department of Physics, Kent State University, Kent OH 44242 USA}
\affiliation{Physics Division, Argonne National Laboratory, Argonne, IL 60439 USA}

\author{Ian C. Clo\"et}
\affiliation{Physics Division, Argonne National Laboratory, Argonne, IL 60439 USA}

\author{Peter~C.~Tandy}
\affiliation{Center for Nuclear Research, Department of Physics, Kent State University, Kent OH 44242 USA}
\affiliation{CSSM, Department of Physics, University of Adelaide, Adelaide SA 5005, Australia}


\begin{abstract}
The leading-twist parton distribution functions of the pion and kaon are calculated for the first time using a rainbow-ladder truncation of QCD's Dyson-Schwinger equations (DSEs) that self-consistently sums all planar diagrams. The non-perturbative gluon dressing of the quarks is thereby correctly accounted for, which in practice means solving the inhomogeneous Bethe-Salpeter equation (BSE) for the quark operator that defines the spin-independent quark distribution functions. An immediate consequence of using this dressed vertex is that gluons carry 35\% of the pion's and 30\% of the kaon's light-cone momentum, with the remaining momentum carried by the quarks. The scale associated with these DSE results is $\mu_0 = 0.78\,$GeV. The gluon effects generated by the inhomogeneous BSE are inherently non-perturbative and cannot be mimicked by the perturbative QCD evolution equations. A key consequence of this gluon dressing is that the valence quarks have reduced support at low-to-intermediate $x$, where the gluons dominate, and increased support at large $x$. As a result, our DSE calculation of the pion's valence quark distribution is in excellent agreement with the Conway~{\it et al.} pion-induced Drell-Yan data, but nevertheless exhibits the $q_\pi(x) \simeq (1-x)^2$ behavior as $x\to 1$ predicted by perturbative QCD.
\end{abstract}

\maketitle
\looseness=-1
The Standard Model supports only one stable hadron -- the proton -- as such this quark-gluon bound state has received the bulk of experimental and theoretical study in the context of Quantum Chromodynamics (QCD). However, the octet of Goldstone bosons -- the pions, kaons and eta -- play a unique role in QCD, as they are associated with dynamical chiral symmetry breaking (DCSB) and would be massless in the chiral limit (vanishing current quark masses). In addressing the key challenges posed by QCD, it is therefore imperative that the quark-gluon structure of QCD's Goldstone bosons be explored.

Such studies are made more important because in the neighborhood of the chiral limit soft-pion theorems provide a number of exact results for the properties for QCD's Goldstone bosons~\cite{GellMann:1968rz,Goldberger:1958tr,Polyakov:1998ze}. Goldstone bosons are also amenable to analyses within perturbative QCD in certain kinematic limits, such as the large $Q^2$ behavior of the electromagnetic elastic and transition form factors~\cite{Efremov:1978rn,Farrar:1979aw,Lepage:1979zb,Lepage:1980fj}. The domain of validity of such results can be explored through a comparison with experimental data, and further study using non-perturbative methods that maintain a close connection to QCD~\cite{Cloet:2013tta,Chang:2013pq,Chang:2013nia}. In this way the inner workings of QCD and the strong interaction can be further revealed. 

A prominent example is the pion's parton distribution functions (PDFs) and their behavior near $x=1$. A longstanding prediction, first based on model field theories and perturbative QCD, and later made rigorous within QCD factorization, is that the pion's quark distribution function behaves as $q_\pi(x) \simeq (1-x)^2$~\cite{Ezawa:1974wm,Farrar:1975yb,Soper:1976jc,Berger:1979du,Yuan:2003fs,Ji:2004hz} and the gluon distribution function as $g_\pi(x) \simeq (1-x)^3$~\cite{Close:1977qx,Sivers:1982wk} in the limit $x\to 1$. The pion's quark distribution has been measured over the domain $0.21 \leqslant x \leqslant 0.99$ in the E615 experiment~\cite{Conway:1989fs} using the pion-induced Drell-Yan reaction $\pi^- N \rightarrow \mu^+ \mu^- X$, and data 
suggests that the pion's quark distribution behaves as $q_\pi(x) \simeq (1-x)$ as $x \to 1$. This contradiction between QCD theory and experiment has been an enduring puzzle in hadron physics~\cite{Wijesooriya:2005ir,Holt:2010vj,Aicher:2010cb}.

Calculations of the pion's valence quark distribution using QCD's Dyson-Schwinger equations (DSEs)~\cite{Cloet:2013jya,Eichmann:2016yit} were first performed in Ref.~\cite{Hecht:2000xa}, with a number of subsequent calculations of various degrees of sophistication~\cite{Nguyen:2011jy,Chang:2014lva,Chang:2014gga,Chen:2016sno,Shi:2018mcb}. These DSE calculations were consistent with the QCD theory expectation as $x \to 1$ and in strong contrast to the E615 data at large $x$. The origin of the agreement between the DSE results and perturbative QCD is that the ultraviolet behavior of the $\bar{q}q$ interaction is dominated by one-gluon exchange in both cases.

\looseness=-1
The apparent disagreement between the E615 data and predictions from QCD theory and the DSEs was highlighted in Ref.~\cite{Holt:2010vj}, which motivated another analysis of the E615 data by Aicher {\it et al.} (ASV)~\cite{Aicher:2010cb} that for first time included soft-gluon summation. This analysis found good agreement with the DSE results for $q_\pi(x)$ and is consistent with perturbative QCD for $q_\pi(x)$ as $x \to 1$. However, the gluon PDF at the initial scale in ASV behaves as $g_\pi(x) \simeq (1-x)^{1.3}$ as $x\to 1$, and therefore disagrees with the perturbative QCD expectation. The DSE results also lacked a self-consistent treatment of the quark and gluon contributions to the pion PDFs, and disagree with the most recent global analysis by the JAM Collaboration~\cite{Barry:2018ort}, which indicates a preference for $q_\pi(x) \simeq (1-x)$ as $x \to 1$ but does not include soft-gluon resummation. Nevertheless, JAM finds $g_\pi(x)/q_\pi(x) = 0$ as $x \to 1$, consistent with perturbative QCD.

In this work we revisit this puzzle using the DSEs and make a significant improvement over previous calculations by treating the quark and gluon contributions to the PDF self-consistently. This is achieved by solving the inhomogeneous Bethe-Salpeter equation (BSE) for the dressed quark operator that defines the spin-independent quark distribution functions. This self-consistent gluon dressing maintains the $q_\pi(x) \simeq (1-x)^2$ behavior predicted by perturbative QCD but pushes its onset to much larger $x$. The new DSE result is consistent with the E615 data from Conway {\it et al.}~\cite{Conway:1989fs}, perturbative QCD expectations for $q_\pi(x)$ and $g_\pi(x)$ as $x\to 1$~\cite{Farrar:1975yb,Berger:1979du,Close:1977qx,Sivers:1982wk}, and the recent JAM Collaboration analysis~\cite{Barry:2018ort} for $x < 1$.

These calculations illustrate two important points: 1) The perturbative QCD predictions 
emphasize the derivative at $x=1$ and need not influence the physically accessible domain;
2) A robust QCD analysis of the large-$x$ behavior of the pion's quark distribution should treat the gluon distribution self-consistently, which should imply that the gluon PDF is more suppressed at large $x$ than the quark PDF, because the large-$x$ quarks are the source of the large-$x$ gluons~\cite{Close:1977qx,Sivers:1982wk}. 

The leading-twist pion and kaon quark distribution functions, which for a quark of flavor $q$ are defined by~\cite{Jaffe:1983hp,Ellis:1991qj,Jaffe:1996zw,Diehl:2003ny}:\footnote{The Wilson line in Eq.~\eqref{pdfdef} enforces color gauge invariance and in light-cone gauge ($n\cdot A=0$) becomes unity~\cite{Jaffe:1996zw}. The DSE approach used here is formulated in Landau gauge, so the gauge link can in principle make a non-trivial contribution to the PDF. However, we leave the challenge of a DSE treatment for the Wilson line to future work.}
\begin{align}
\hs*{-2mm}q(x) =\!\! \int\! \frac{d \lambda}{4 \pi}\, e^{-i\,xP \cdot n\,\lambda}\,
\langle P | \widebar{\psi}_q(\lambda n)\,\sh{n}\,W(\lambda,n \cdot A)\,\psi_q(0) | P \rangle_c,
\label{pdfdef} 
\end{align}
where $n$ is a light-like 4-vector ($n^2=0$)~\cite{Jaffe:1996zw,Diehl:2003ny,Collins:2011zzd}, are evaluated here using QCD's DSEs in the rainbow-ladder truncation defined in Ref.~\cite{Qin:2011dd}. The DSEs have proven to be a powerful tool with which to study hadron structure~\cite{Cloet:2013jya,Eichmann:2016yit}, with particular success in predicting the properties of the Goldstone bosons~\cite{Maris:2000sk,Maris:2002mz,Chang:2013pq,Chang:2013nia} since the DSE framework encapsulates many aspects of DCSB and quark-gluon confinement in QCD~\cite{Cloet:2013jya,Bednar:2018htv}. 

\begin{figure}[tbp]
\centering\includegraphics[width=0.985\columnwidth]{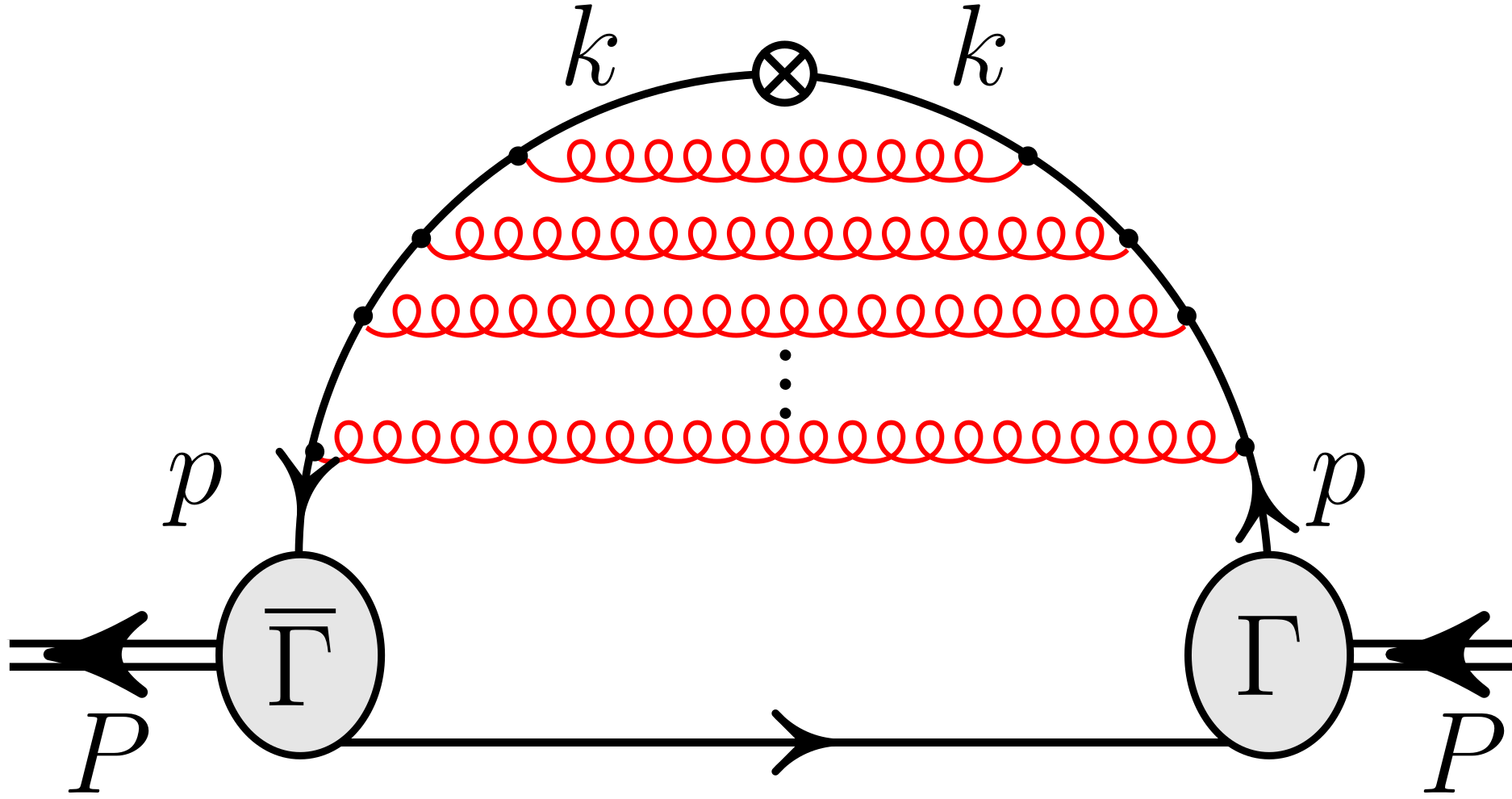}
\caption{Diagram that gives the quark distribution in a pion or kaon, where all quark and gluon propagators are dressed. The elementary operator insertion is given by $\otimes = i\,Z_2\,\sh{n}\,\delta\!\lf(x - k \cdot n/P \cdot n\rg)$.}
\label{fig:modelscalepion}  
\end{figure}

Within a rainbow-ladder truncation to QCD's DSEs the quark distribution functions in a meson $M$ can be represented by the diagram in Fig.~\ref{fig:modelscalepion}, and expressed as
\begin{align}
q_M(x) &= \frac{1}{2\,P\cdot n}\ {\rm Tr} \int\! \frac{d^4p}{(2\pi)^4}\ \widebar{\Gamma}_M(p,P)\,S(p)\,\Gamma_q(x,p,n) \no \\
&\hs*{32mm} \times
S(p)\,\Gamma_M(p,P)\,S(p-P),
\label{eq:dsepdfs}
\end{align}
where the trace is over Dirac, color, and flavor indices. The dressed quark propagator $S(p)$ is obtained by solving the gap equation~\cite{Qin:2011dd,Cloet:2013jya,Eichmann:2016yit} and the bound-state amplitude $\Gamma_M(p,P)$ is the solution to the homogeneous BSE~\cite{Maris:1997tm,Maris:2000sk,Qin:2011dd}, which in the rainbow-ladder truncation sums an infinite number of ladder gluon exchanges between the quark and antiquark. In Fig.~\ref{fig:modelscalepion} these gluons have been absorbed into $\Gamma_M(p,P)$ and $\widebar{\Gamma}_M(p,P)$. The vertex $\Gamma_q(x,p,n)$ represents the infinite sum of exchanged dressed-gluons in Fig.~\ref{fig:modelscalepion} and satisfies the inhomogeneous BSE
\begin{align}
&\Gamma_q(x,p,n) =  iZ_2\,\sh{n}\,\delta\!\lf(x - \frac{p \cdot n}{P \cdot n}\rg)\nonumber \\
&\hs*{8mm}
- \int\frac{d^4\ell}{(2\pi)^4}\ \gamma_\mu\,\mathcal{K}_{\mu \nu}(p-\ell)\,S(\ell)\,\Gamma_q(x,\ell,n)\,S(\ell)\,\gamma_\nu, 
\label{eq:Gamma}
\end{align}
where $Z_2$ is the quark wave function renormalization. The rainbow-ladder BSE kernel has the form $\mathcal{K}_{\mu \nu}(q) = Z_2^2\,\vect{G}(q^2)\,D^0_{\mu\nu}(q)$, with $D^0_{\mu\nu}(q)$ the bare gluon propagator in Landau gauge, and $\vect{G}(q^2)$ the effective running coupling whose infrared behavior is governed by a single parameter~\cite{Qin:2011dd} and in the ultraviolet is one-loop renormalized QCD.

Physical insight into Eq.~\eqref{eq:dsepdfs} can be obtained by introducing  $\ident = \iint dydz\,\delta\!\lf(y - \frac{p\cdot n}{P\cdot n}\rg)\delta\big(z - \frac{k\cdot n}{p\cdot n}\big)$, and in that context obtain
\begin{align}
\hs*{-2.0mm}\Gamma_q(x,p,n) = \iint dy\,dz\, \delta(x-yz)\,\delta\!\lf(y - \frac{p \cdot n}{P \cdot n}\rg)\Lambda_q(z,p,n),
\label{eq:Gamma2}
\end{align}
where $z = k\cdot n/p\cdot n = x P\cdot n/p\cdot n = x/y$ is the light-cone momentum fraction of the ``active'' quark relative to the host dressed quark, and $y$ is the host dressed quark light-cone momentum fraction relative to the meson. The hadron-independent vertex $\Lambda_q(z,p,n)$ is represented by the inhomogeneous BSE:
\begin{align}
&\Lambda_q(z,p,n) = iZ_2\,\sh{n}\,\delta(1 - z) - \iint du\,dw\,\delta(z-uw)\int\! \frac{d^4\ell}{(2\pi)^4} \nonumber \\
&\hs*{1mm} \times
\delta\!\lf(w - \frac{\ell\cdot n}{p\cdot n}\rg)\ \gamma_\mu\,\mathcal{K}_{\mu \nu}(p-\ell)\, S(\ell)\, \Lambda_q(u,\ell,n)\, S(\ell)\, \gamma_\nu,
\label{eq:Lambda}
\end{align}
which has a solution of the form
\begin{align}
\Lambda_q(z,p,n) &= i\sh{n}\,\delta(1 - z) \no \\
&\hs*{-5mm}
+ i\sh{n}\,f_1^q(z,p^2) + n \cdot p\lf[i\sh{p}\,f_2^q(z,p^2) + f_3^q(z,p^2)\rg].
\label{eq:Lambda2}
\end{align}
The functions $f_i^q(z,p^2)$ can be interpreted as quark distributions for an active quark inside a dressed quark with virtuality $p^2$.

\looseness=-1
The key advance in this work is a rigorous treatment of the dressed quark operator $\Lambda_q(z,p,n)$ in a rainbow-ladder truncation to QCD's DSEs, which allows for a self-consistent distinction between the momentum carried by quarks and gluons. Two approximations to Eq.~\eqref{eq:Lambda} are common: the simplest is to ignore the gluon dressing which is natural in Nambu--Jona-Lasinio type models~\cite{Cloet:2005pp,Cloet:2007em,Ninomiya:2017ggn}, another, adopted in previous DSE-based approaches, is the \textit{Ward-identity ansatz} (WIA):
\begin{align}
\Lambda_q(z,p,n) \to \Lambda_q^{\rm WIA}(z,p,n) = \delta\lf(1 - z\rg)\ 
n^\mu\, \frac{\partial}{\partial p^\mu}\,S_q^{-1} (p).
\label{eq:WIA}
\end{align}
This ansatz approximates the true light-cone momentum fraction $z = k\cdot n/p\cdot n$ by $z = 1$, which in rainbow-ladder truncation is exact for the zeroth moment, but breaks down for any higher moment and therefore does not distribute momentum in a physical way between quarks and gluons. To date all DSE studies have employed the WIA.

\begin{figure}[tbp]
\centering\includegraphics[width=\columnwidth,height=160mm]{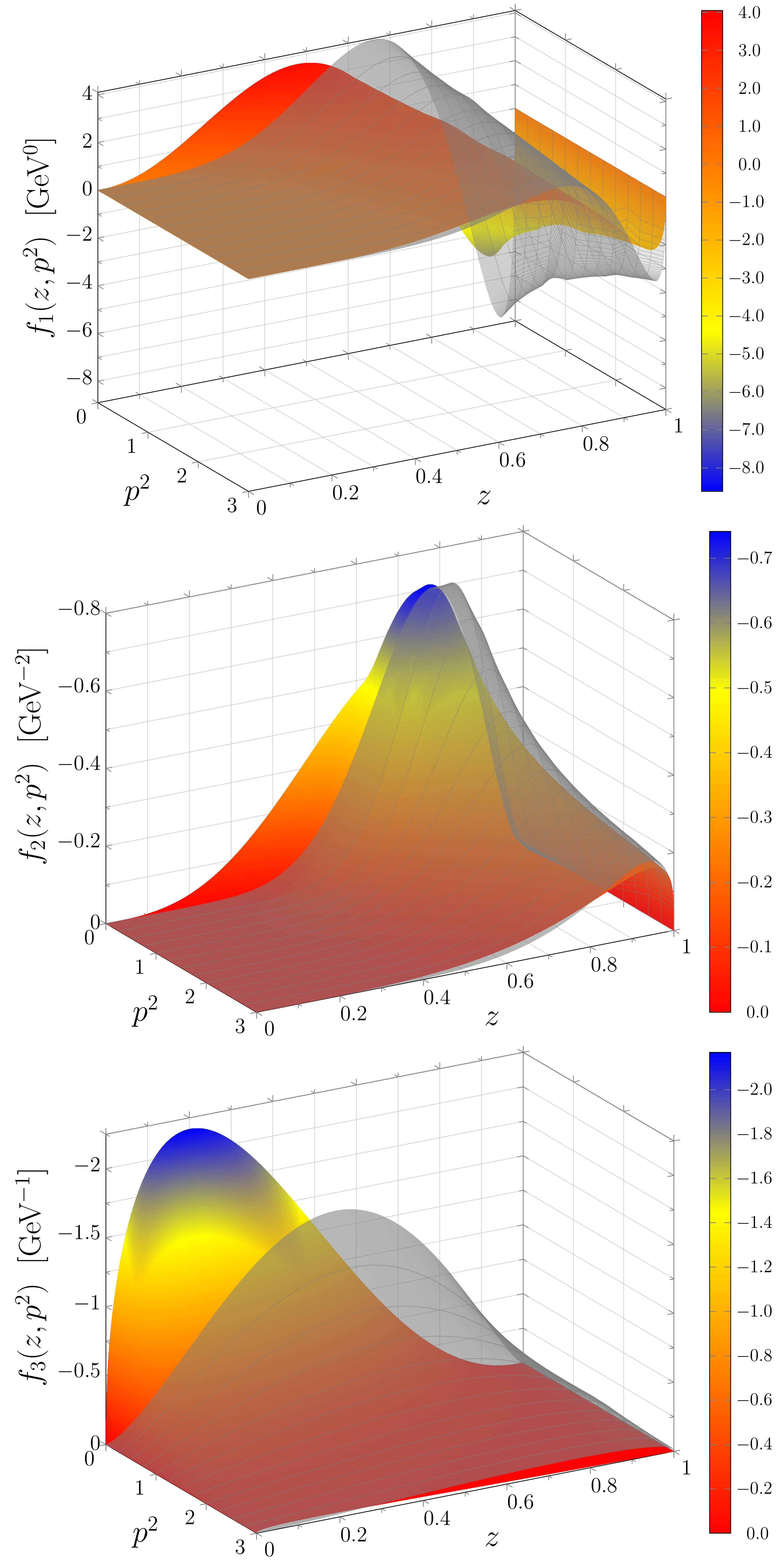}
\caption{Results for $f_i^q(z,p^2)$ from Eq.~\eqref{eq:Lambda2}, which can be interpreted as a PDF for the active quark in a parent dressed quark of virtuality $p^2$. The colored surfaces represent results for $u/d$ quarks, while the gray surface depicts $s$ quark results relevant to the kaon.}
\label{fig:verts}  
\end{figure}

The DSEs are formulated in Euclidean space and therefore a direct evaluation of Eq.~\eqref{eq:dsepdfs} to obtain the pion and kaon PDFs is challenging because of non-analytic structures in the complex plane. This can be alleviated by first calculating a finite number of moments -- defined by $\langle x^n \rangle_M^q = \int_0^1\,dx\,x^n\,q_M(x)$ -- and then reconstructing the PDFs from these moments. As a first step we must solve Eq.~\eqref{eq:Lambda} -- using the moment method -- from which we can reconstruct the functions $f_i^q(z,p^2)$. 

Results for $f_i^q(z,p^2)$ are given in Fig.~\ref{fig:verts} for $u/d$ and $s$ (active) quark PDFs in a parent dressed quark of the same flavor -- there is no flavor-mixing in rainbow-ladder truncation. These results are obtained by solving for a large number of $p^2$-dependent moments of Eq.~\eqref{eq:Lambda} that are then accurately fitted by a function of the form 
$f(z,p^2) = N(p^2)\, z^{\alpha(p^2)}\, (1-z)^{\,\beta(p^2)}\lf[1 + c(p^2)\,\sqrt{z}\rg]$.
Our results for $f_i^q(z,p^2)$ show significant support over the entire $z$ domain, which is a consequence of the non-perturbative gluon dressing of the elementary quark operator and a direct indication that gluons carry significant momentum. This can be contrasted with the elementary vertex where the functions $f_i^q(z,p^2)$ vanish and the WIA where all strength is concentrated at $z=1$. The $s$ quark distributions have support concentrated closer to $z=1$ relative to the light quarks, which indicates that the active $s$ quark tends to carry a larger light-cone momentum fraction of the parent and that gluon dressing effects are suppressed.

For each $f_i^q(z,p^2)$ we find that as the dressed quark virtuality increases gluon dressing effects diminish, and since the renormalization condition for the dressed quark propagator is $S_q^{-1}(p)\big|_{p^2=\mu^2} = i\sh{p} + m_q$, where $m_q$ is the renormalized current quark mass of flavor $q$, the functions $f_i^q(z,p^2)$ vanish when $p^2=\mu^2$. The results in Fig.~\ref{fig:verts} have the renormalization scale set to $\mu = 2\,$GeV, and $m_u = m_d$ which implies $f_i^u(z,p^2) = f_i^d(z,p^2)$. Importantly, these PDF results for an active quark in a parent dressed quark of virtuality $p^2$ are universal for any DSE calculation of hadron quark distributions, and are not specific to pion and kaon PDFs which is the focus here. In a calculation of hadron PDFs the quantities $f_i^q(z,p^2)$ appear with an integration over the dressed quark virtuality.

\begin{table}[tbp]
\addtolength{\tabcolsep}{7.0pt}
\addtolength{\extrarowheight}{1.8pt}
\begin{tabular}{l|ccccc}
\hline\hline
PDF & $Q$~(GeV) & $\lf< x \rg>$ & $\lf< x^2 \rg>$ & $\lf< x^3 \rg>$ & $ \lf< x^4 \rg> $  \\
\hline
$u_{\pi}$       &  0.78  &  0.323  &  0.167  &  0.109  &  0.083 \\
$u_{K}$         &  0.78  &  0.297  &  0.148  &  0.092  &  0.065 \\
$\bar{s}_{K}$   &  0.78  &  0.402  &  0.221  &  0.143  &  0.101 \\
\hline
$u_{\pi}$       & 1.3\ph{0}    & 0.268   & 0.125   & 0.076   & 0.054 \\
JAM  & 1.3\ph{0} & 0.268 & 0.127 & 0.074 & 0.048 \\
ASV  & 1.3\ph{0} & 0.247 & 0.106 & 0.055 & 0.033 \\
LQCD & 2.0\ph{0} & 0.27\ph{0}  & 0.13\ph{0} & 0.074 & ---\\
\hline
\end{tabular}
\caption{{\it Top-panel:} DSE results at the model scale for low moments of the $u$ quark distribution in the $\pi^+$, and the $u$ and $\bar{s}$ quark distributions in the $K^+$. {\it Bottom-panel:} Comparisons between our pion DSE results and analyses from JAM~\cite{Barry:2018ort} and ASV~\cite{Aicher:2010cb}, and lattice QCD~\cite{Brommel:2006zz}.}
\label{tab:moments}
\end{table}

Using our results for $\Lambda_q(z,p,n)$  we can now numerically evaluate a finite number of moments for the pion and kaon quark distributions defined in Eq.~\eqref{eq:dsepdfs}. Our DSE framework conserves baryon number, and therefore $\langle x^0 \rangle_M^q = 1$ for each valence quark $q$. In Tab.~\ref{tab:moments} we present results for four non-trivial moments. Since the DSE framework developed here is self-consistent and therefore momentum-conserving, the results in Tab.~\ref{tab:moments} and the momentum sum rule $\langle x \rangle_M^u + \langle x \rangle_M^{\bar{u}} +  \ldots + \langle x \rangle_M^g = 1$ then immediately imply that gluons carry 35\% of the pion's and 30\% of the kaon's light-cone momentum at the DSE scale, with the reminder carried by the quarks. The DSE scale is found to be $\mu_0=0.78\,$GeV by matching the DSE quark momentum fraction to that found by JAM~\cite{Barry:2018ort}, and it is clear from Tab.~\ref{tab:moments} that the heavier $s$ quark suppresses gluon momentum fractions. In the context of the triangle diagram of Fig.~\ref{fig:modelscalepion}, and in the rainbow-ladder truncation, there are two processes that allow gluons to carry momentum: 1) the gluon dressing of the quark operator, and 2) the gluons that are exchanged between the quark and antiquark pair to form the bound state. Comparison with PDF results obtained using a bare vertex and the WIA indicate that the gluons that dress the quark operator carry the bulk of the gluon momentum in the bound state.

\begin{table}[tbp]
\addtolength{\tabcolsep}{3.6pt}
\addtolength{\extrarowheight}{1.0pt}
\begin{tabular}{l|ccc}
\hline\hline
PDF  & $N$ & $\alpha_i$ & $c_i$ \\
\hline
$u_\pi$     & $9.567$ & $\lf[1.702,\,\ph{-}7.857,\,34.19\rg]$ & $\lf[3.688,\,21.17,\,\ph{-}2.699\rg]$  \\
$u_K$       & $12.16$ & $\lf[4.344,\,    -4.516,\,12.23\rg]$ & $\lf[0.258,\,4.249,\,-2.883\rg]$  \\
$\bar{s}_K$ & $25.37$ & $\lf[5.497,\,    -4.425,\,14.46\rg]$ & $\lf[0.324,\,4.768,\,\ph{-}1.099\rg]$  \\
\hline
\end{tabular}
\caption{Fit parameters for Eq.~\eqref{eq:fitform} that give the valence quark distributions at the model scale of $\mu_0 = 0.78\,$GeV.}
\label{tab:pdfparams}
\end{table}

\begin{figure}[tbp]
\centering\includegraphics[width=\columnwidth,height=58mm]{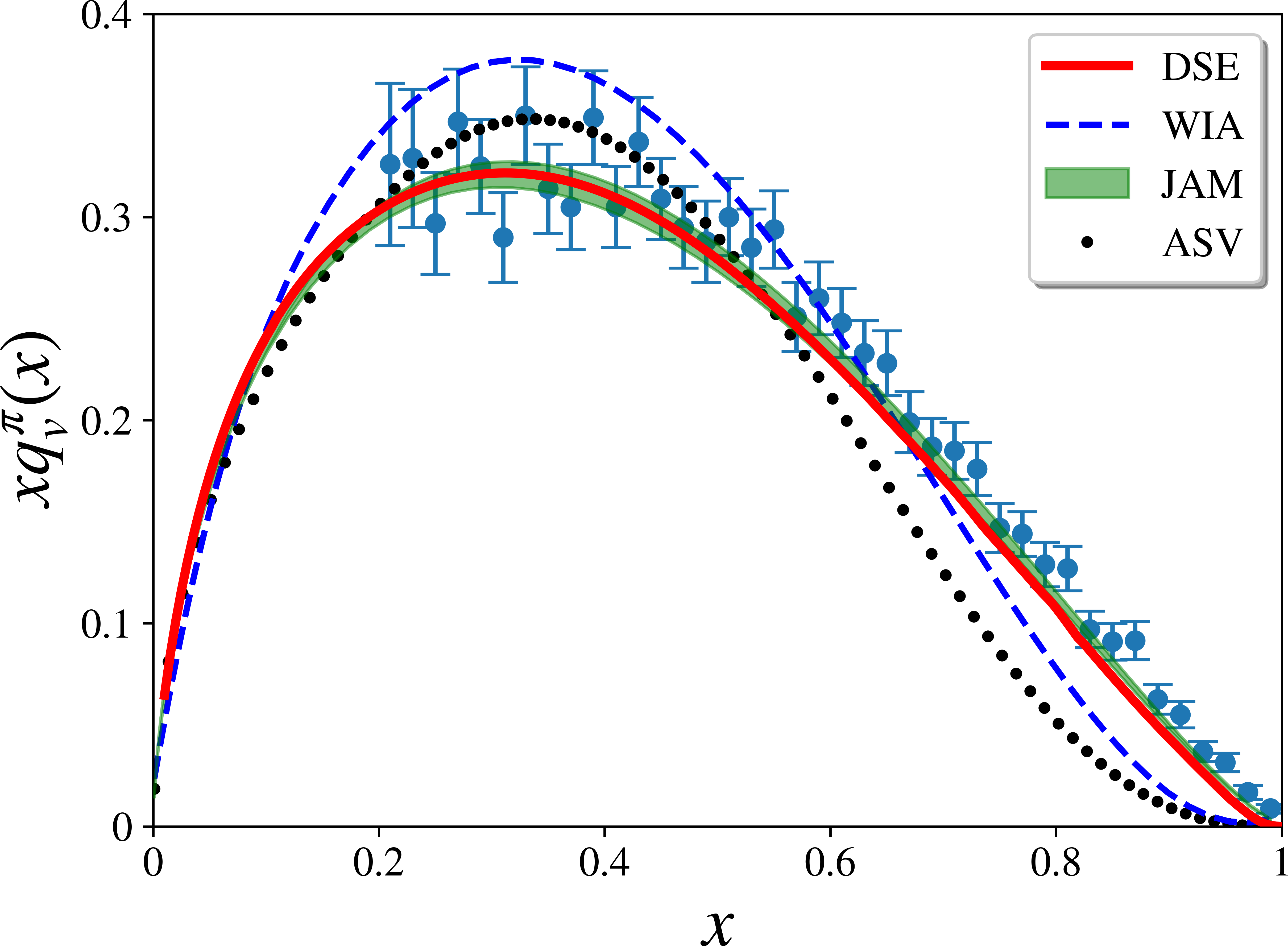} \\[0.5em]
\centering\includegraphics[width=\columnwidth,height=58mm]{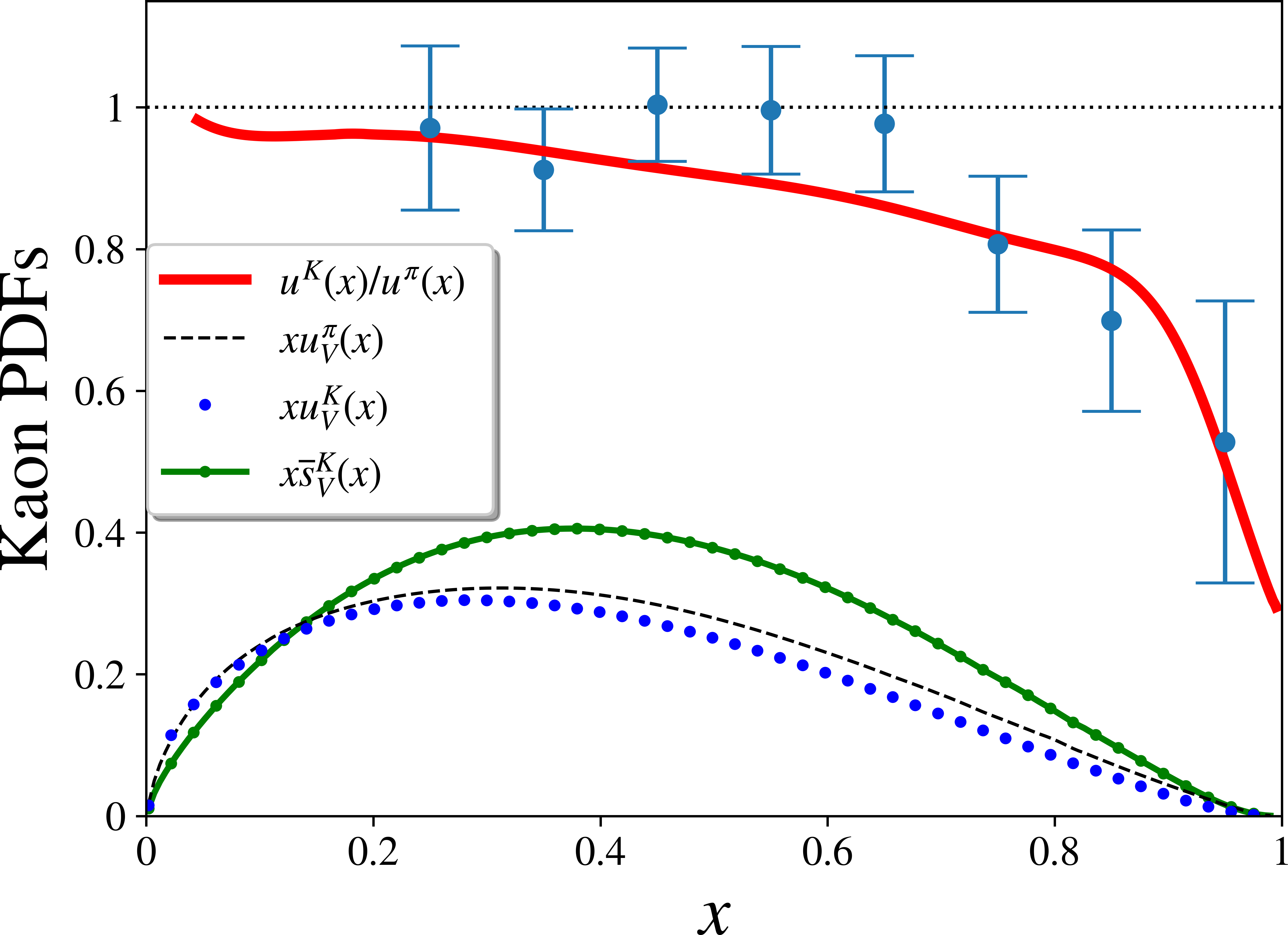}
\caption{{\it Top panel:} Solid line is our DSE result for the pion's valence quark distribution; green band is the JAM analysis~\cite{Barry:2018ort}; dashed curve is the DSE result from Ref.~\cite{Nguyen:2011jy} which uses the WIA; and the dotted curve is the NLO soft-gluon-resummation analysis from ASV~\cite{Aicher:2010cb}.  All curves are at the scale $Q = 5.2\,$GeV and, with the exception of JAM, have a $x\to 1$ exponent consistent with theory expectations.
{\it Bottom panel:} Contrast between our DSE results for the pion and kaon PDFs, and $u_K(x)/u_\pi(x)$ which is compared to data from Ref.~\cite{Badier:1980jq}.
}
\label{fig:pionPDF}  
\end{figure}

To re-construct the PDFs from the moments we fit them to
\begin{align}
q(x) &=  N\,x^{\alpha_1} (1 - x)^2\lf(1 + c_1 x^{\alpha_2} + c_2 x^{\alpha_3}\rg) + c_3 (1-x)^4, 
\label{eq:fitform}
\end{align}
with the resulting fit parameters provided in Tab.~\ref{tab:pdfparams} for the pion and kaon valence PDFs. The DSE framework has been shown to give an exponent $q(x) \simeq (1-x)^2$ as $x \to 1$ for each quark distribution~\cite{Hecht:2000xa,Nguyen:2011jy,Chang:2014lva}; we therefore impose this constraint in Eq.~\eqref{eq:fitform}. DSE results for the pion and kaon quark distributions are presented in Fig.~\ref{fig:pionPDF} at a scale of $Q = 5.2\,$GeV. Our DSE result for the pion is in excellent agreement with the pion-induced Drell-Yan measurements of Conway {\it et al.}~\cite{Conway:1989fs} over the entire $x$ domain of the data, as well as the recent JAM analysis~\cite{Barry:2018ort}. The only kaon data is for the ratio $u_K(x)/u_\pi(x)$~\cite{Badier:1980jq}, and again we find excellent agreement.\footnote{To perform the singlet DGLAP evolution needed in this case we take a gluon distribution of the form $g(x) = N_g\,x^{-1}(1-x)^3$~\cite{Brodsky:1989db,Hoyer:1997rh} and use our results for the gluon momentum fraction to constraint $N_g$.}

Fig.~\ref{fig:pionPDF} contrasts our DSE result for $u_v^\pi(x)$ with the rainbow-ladder DSE calculation from Ref.~\cite{Nguyen:2011jy} that used the WIA. The self-consistent treatment of the gluon contributions to the PDFs has a dramatic impact on the quark distributions over the entire valence region, except near $x=1$. The differences between our DSE result and earlier results using the WIA can be understood as follows: the non-perturbative gluons that dress the quark operator that defines the PDFs dominate at low-to-intermediate $x$, and in this domain carry significant momentum, thereby reducing support for the quark distributions in this region. However, the valence quark PDFs must satisfy the baryon number sum rule, which necessitates increased support at large $x$ where the gluons play less of a role. This shift in support for the quark PDFs from the gluon dressing is inherently non-perturbative, because here the quark-gluon splitting functions are dressed, and therefore cannot be mimicked by the perturbative gluons introduced by the QCD evolution equations. 

In our DSE calculation the large-$x$ quarks are the source of the large-$x$ gluons, which implies $g(x)/q(x) \to 0$ as $x\to 1$ at the initial DSE scale, in agreement expectations from perturbative QCD~\cite{Close:1977qx,Sivers:1982wk}. This is in contrast to the ASV analysis~\cite{Aicher:2010cb} where $g_\pi(x)/u_\pi(x) \to \infty$ as $x\to 1$ at the initial scale of the PDFs. This observation may be the source of the differences between the ASV analysis illustrated in Fig.~\ref{fig:pionPDF} and our DSE result. This illustrates the importance of a self-consistent treatment of both the quark and gluon contributions in any PDF analysis or calculation.

The pion and kaon PDF results presented here include for the first time a correct treatment of the non-perturbative gluon dressing of the quark operator that defines the spin-independent quark distributions within a rainbow-ladder truncation to QCD's DSEs. With this framework it is straightforward to correctly distinguish between quark and gluon contributions to PDFs, which is shown to have a dramatic impact on quark PDFs over the entire valence region. An immediate consequence of this non-perturbative gluon dressing is that gluons carry 35\% of the pion's and $30\%$ of the kaon's light-cone momentum at the initial scale of the DSE calculations. Our results for the pion and kaon PDFs are in excellent agreement with available data, and agree with the perturbative QCD expectation as $x \to 1$. These results demonstrate that a self-consistent analysis of both the quark and gluon PDFs is essential, and that more data on both distributions in the pion and kaon is needed, e.g., from the proposed electron-ion collider.

\begin{acknowledgments}
KB thanks Chao Shi for several beneficial conversations. This work was supported by the U.S. Department of Energy, Office of Science, Office of Nuclear Physics, contract no.~DE-AC02-06CH11357; by the National Science Foundation, grant no.~NSF-PHY1516138; the Laboratory Directed Research and Development (LDRD) funding from Argonne National Laboratory, project no.~2016-098-N0 and project no.~2017-058-N0; and the DOE Office of Science Graduate Student Research (SCGSR) Program.
\end{acknowledgments}


%

\end{document}